\documentclass[twocolumn,pre,floats,aps,amsmath,amssymb,nofootinbib]{revtex4-1}
\usepackage{graphicx}
\usepackage{bm}
\usepackage{physics}

\begin{document}

\title{Relativistic Implications for 
Physical Copies of Conscious States}
\author{Andrew Knight}
\affiliation{aknight@alum.mit.edu}
\date{\today}

\begin{abstract}
The possibility of algorithmic consciousness depends on the assumption that conscious states can be copied or repeated by sufficiently duplicating their underlying physical states, leading to a variety of paradoxes, including the problems of duplication, teleportation, simulation, self-location, the Boltzmann brain, and Wigner's Friend.  In an effort to further elucidate the physical nature of consciousness, I challenge these assumptions by analyzing the implications of special relativity on evolutions of identical copies of a mental state, particularly the divergence of these evolutions due to quantum fluctuations.  By assuming the supervenience of a conscious state on some sufficient underlying physical state, I show that the existence of two or more instances, whether spacelike or timelike, of the same conscious state leads to a logical contradiction, ultimately refuting the assumption that a conscious state can be physically reset to an earlier state or duplicated by any physical means.  Several explanatory hypotheses and implications are addressed, particularly the relationships between consciousness, locality, physical irreversibility, and quantum no-cloning.
\end{abstract}

\maketitle

\section{Introduction}
\label{sec:intro}

Either it is physically possible to copy or repeat a conscious state or it is not.  Both possibilities have profound implications for physics, biology, computer science, and philosophy.  

Whether or not consciousness can be copied is fertile ground for a multitude of troubling, if not fascinating, thought experiments.  There's the duplication problem \cite{Penrose}: imagine we can teleport a traveler to another planet by creating ``a precise duplicate of the traveler, together with all his memories, his intentions, his hopes, and his deepest feelings," but then we decide not to destroy the original copy?  ``Would his `awareness' be in two places at once?"  There's the simulation problem \cite{Bostrom}: if conscious awareness can be uploaded onto a computer, then how do we know we aren't simulated minds in simulated universes?  In fact, because simulated universes are much less expensive, in terms of matter and energy, than actual universes, then if consciousness can be duplicated, we are almost certainly one of vast numbers of simulated copies.  There's the problem of self-location \cite{Elga}: if a psychopath tells you that he has created an exact physical copy of you and will torture it unless you pay a hefty ransom, then you should pay the ransom unless you are absolutely sure that \textit{you} aren't the copy.  There's the problem of the Boltzmann brain: if consciousness is just the result of atoms in a brain, what's to prevent a set of physically identical atoms, somewhere in the universe, from accidentally coming together in just the right way to create your brain?  And what would that feel like?

Notice that each of these problems is a direct consequence of the \textit{copiability} or \textit{repeatability} of conscious states.\footnote{Throughout this paper, I'll treat copiability and repeatability of conscious states as essentially interchangeable because repeating a conscious state (or resetting it to an earlier state) is akin to copying it at a later point in spacetime.}   If it turns out, for whatever reason, that conscious states cannot be copied, then these and other problems disappear.  Whether each of the above scenarios is actually possible is an empirical question, and given the rate that technology is advancing, it may be just a matter of time before each is tested.  There are related problems that may never be testable such as, ``Will a computer ever become conscious -- and how would we know?"  After all, there is no consensus on how to measure the existence or level of consciousness in an entity.  Some say that consciousness depends on the ability to pass a hypothetical ``Turing test."  Some say it depends on the level of complexity in neural networks.  Some say it depends on certain activity in the brain.  So how can we possibly learn anything about the nature of consciousness if it depends on a definition?  

It is easy to get bogged down in the meaning of ``conscious" and lose sight of the big picture.  It's as if the question, ``Will a ball near Earth experience a force of gravity?" has been preempted by, ``What do you mean by `ball'?"  So let me rephrase: ``If I am near Earth, will \textit{I} experience a force of gravity?"  My goal is to derive, if possible, an objectively correct prediction to several questions such as: will it ever be possible to teleport a copy of myself to another planet?  Will it ever be possible to upload my conscious awareness onto a computer so that I can outlive my physical death?  Will it ever be possible, perhaps in millions of years, for a collection of atoms somewhere in the universe to accidentally come together in just the right way to create my conscious awareness?  Will it be possible for me to experience these things?  The answer is yes only if my conscious states are fundamentally such that they could be copied or repeated.  Are they?

First, let's consider the question, ``Will it ever be possible to upload me to a digital computer?"  This a scientific question if it can be phrased as a falsifiable hypothesis.  ``It is impossible to upload me to a digital computer" would be falsified, for example, by my being conscious and aware while possessing a body consisting essentially of on-off switches.  But what do I mean by ``me"?  I mean my identity -- but what is that?  Questions like this lead many to reject from the realm of scientific inquiry anything involving identity or consciousness.  But that's a mistake because science is about making predictions about our own experiences, and without a consistent sense of identity, \textit{whose} experiences are we predicting?

In this paper, I will assume it is possible to copy conscious states and show how this assumption, in conjunction with special relativity, leads to a contradiction.  In doing so, I will rely heavily on the extent to which consciousness and identity supervene on an underlying physical state.  For this reason, I will begin with a discussion of identity and why objections based on identity are inadequate to counter the conclusions of this paper.

\section{Questioning Identity}
\label{sec:identity}

Let's consider what it means to upload me to a digital computer.  I think it means two things.  First, that it's me, that I experience it.  (Let's call this \textit{existence} of identity.)  If a scientist tells me that an exact copy of me has been uploaded to a computer, but I don't experience whatever sensations accompany living inside a computer, I will respond that I have not been uploaded.  ``But I can't tell the difference between you two," the scientist may respond.  ``Yes, but I can," I would reply, ``and you have failed to upload me."  Second, that my identity would persist; that as the computer physically evolves in time according to the laws of nature, the consciousness it creates would continue to be mine.  (Let's call this \textit{persistence} of identity.)  In other words, if the computer begins in some initial physical state $S_1$ that produces me in conscious state $C_1$, and after awhile physically evolves to physical state $S_2$ that happens to produce conscious state $C_2$, that conscious state $C_2$ will still be mine, and I subjectively experience an evolution from conscious state $C_1$ to $C_2$.  And if, due to a random quantum event, the computer were to evolve to physical state $S_2^{'}$ (instead of $S_2$) that happens to produce $C_2^{'}$ (instead of $C_2$), then conscious state $C_2^{'}$ will still be mine and I subjectively experience an evolution from conscious state $C_1$ to $C_2^{'}$.  I don't expect the scientist to guarantee that I will live forever -- after all, the physical evolution of the computer could simply stop producing consciousness -- but I do expect that as long as the physical evolution of the computer continues to produce conscious states, I will be the person experiencing them, not someone else.  In other words, if a scientist claims to have uploaded me to a digital computer, if I don't subjectively experience (in the form of evolving conscious states) the physical evolution of the computer over time, then that scientist has failed.  Let me combine existence and persistence of identity into one concept: temporal continuity\footnote{This temporal flow is often called a ``stream" of consciousness, but many have argued that consciousness consists of a series of discontinuous blips of awareness that are perceived as continuous.  This distinction is irrelevant to the present analysis.} of identity (or \textit{transtemporal identity}).  Whether there really is such a thing, and whether it matters, will be addressed in the following sections.

In this paper the word \textit{consciousness}\footnote{Just as there is no consensus on the definition of consciousness, there is no single term or phrase to represent it -- others are mentality, sentience, self-awareness, and the subjective experience of qualia.} includes identity and a \textit{conscious state} is a state experienced by someone; that is, a person experiencing a conscious state has an identity and one conscious state cannot be experienced by more than one person.  So if a particular physical state produces a conscious state, and that physical state includes (among other things) light sensors that are absorbing photons having a wavelength of 680nm, then the produced conscious state may correspond to someone's subjective experience of seeing the color red, in which case we can ask \textit{who} is experiencing that conscious state and whether that person is seeing red.

\subsection{Boltzmann Brains and Bad Science}
\label{sec:BB}

As it turns out, the Boltzmann brain problem is much bigger than the weirdness of the occasional human brain fluctuating into and out of existence.  In cosmological models in which Boltzmann brains dominate, not only is any brain much more likely to be a random fluctuation, but any complete person, world, or even galaxy is much more likely to be a random fluctuation.  Are you a Boltzmann brain?  You might think the answer is no because you can see a vast world around you and you seem to have many years' worth of memories that are consistent with a long-lasting, law-abiding physical world.  However, the overwhelming majority of people who would make such claims are, in these models, random fluctuations.  They are ``Boltzmann observers" who have no right to believe that any of their observations are reliable or that their scientific predictions will come to fruition.  

This is a problem.  Science, after all, is about collecting evidence and making predictions based on falsifiable hypotheses.  However, if a model based on science predicts that the foundation of science itself is not reliable -- then what?  Sean Carroll eloquently sums it up \cite{Carroll}: ``it's overwhelmingly likely that everything we think we know about the laws of physics, and the cosmological model that predicts we are likely to be random fluctuations, has randomly fluctuated into our heads."  In other words, if a model that is based on physical evidence in the world around us predicts that that very evidence is an illusion, then we should reject such a model as ``cognitively unstable."  

A similar fate befalls denial of identity because the foundation of science fundamentally depends on transtemporal identity.  ``A ball falls toward Earth at $9.8  m/s^2$" is an hypothesis that, if incorrect, can be falsified through experimentation.  But that requires a scientist to prepare the experiment, then perform it, then collect and analyze data, and then determine whether the hypothesis has been falsified.  This series of events, all of which are necessary in scientific investigation, occurs in a temporal order, and if the scientist did not or could not expect to be the same person throughout, then she would have no reason to believe that she could test falsifiable hypotheses.  She would have no reason to trust science.  Therefore, any theory or model that rejects the temporal continuity of identity should be rejected as rendering science internally inconsistent.  Further, while transtemporal identity is necessary for scientific inquiry, it is not itself subject to scientific scrutiny because no experiment can falsify it; thus, there is no scientific rationale to doubt the existence of transtemporal identity.

\subsection{Fissioning Philosophers}
\label{sec:philos}

Philosophers have had much to say about both the existence and persistence of identity, but their analyses almost exclusively depend on the assumption that conscious states can be copied.  The typical philosophical analysis of identity goes something like this: \textit{Identity is produced by the brain; the brain is just a chuck of matter that can be moved around, copied, replaced, and so forth, at will; therefore, the notion of identity is problematic.}

For instance, Parfit begins his analysis \cite{Parfit} with, ``We suppose that my brain is transplanted into someone else's (brainless) body, and that the resulting person has my character and apparent memories of my life.  [I shall here assume] that the resulting person is me."  He then proceeds, in thought experiment, to slice up the brain, trade brain sections with other people, sever connections between hemispheres, and so forth, to arrive at various conclusions (such as the ``fission" or ``fusion" of identities) that are at odds with transtemporal identity.  Zuboff \cite{Zuboff} also starts with replication of a brain that ``is a precise duplicate of yours in every discriminable respect..." and concludes that so-called fission -- in which ``one subject \textit{can}, in a single next moment, experience two differing non-integrated contents of experience" -- is the necessary outcome.  These ``brain-in-a-vat" style thought experiments have dominated philosophical theories of mind and identity, not to mention science fiction, for decades, and represent the best \textit{prima facie} arguments against transtemporal identity.

However, an argument that relies on premise X cannot be used to disprove an argument for ¬X.  My goal in this paper is to show, among other things, that conscious states cannot be copied.  Because the typical philosophical arguments depend on the in-principle copiability of brains, mental states, etc., then not only are they inadequate to counter the arguments in this paper, they are entirely inapplicable.  In other words, the strongest arguments against transtemporal identity depend on the assumption that conscious states can be copied, but this is the very notion that I am attempting to refute.  Therefore, any notion or theory of identity that depends on the copiability of conscious states (or the copiability of brains in conjunction with the supervenience of consciousness on brains) is simply irrelevant to the present analysis.

\subsection{Universal Beliefs}
\label{sec:beliefs}

Why fight the notion of transtemporal identity in the first place?  After all, it's what we all implicitly believe.  Every decision we make about the future depends on our belief in identity.  Let's say that a travel agent was to offer you the following options:

\begin{itemize}
\item For \$10,000, you will spend a week at a resort in Tahiti;
\item For \$6,000, an exact physical copy\footnote{I have no idea what this phrase means, nor does any other physicist, although many philosophers throw it around casually.  An exact copy in classical phase space is impossible because infinite precision does not exist in the physical world; an exact copy in quantum mechanical Hilbert space is impossible because of quantum no-cloning; and so on.  But for the sake of this section, let's pretend there is such a thing.} of you will be created and will spend a week at a resort in Tahiti while your current body is anesthetized;
\item For \$2,000, a really good copy of you -- one that looks, acts, and talks just like you but is not an exact copy -- will be created and will spend a week at a resort in Tahiti while your current body is anesthetized.
\end{itemize}

In all three cases, the underlying question you'll no doubt ask yourself is, ``What will I experience?"  If you believe that an exact physical copy of you inevitably includes your identity, then you might be tempted to go with the second option, given that the \$4,000 in savings could be enjoyed by \textit{you} at a later date.  And if you believe that a reasonably good copy of you includes your identity, then you might be tempted by the third option.  In other words, your potential choice between the latter two options depends on (your beliefs about) whether your identity can flow to a copy of you, not on whether you have an identity.  When choosing, you are, presumably, not thinking about neural firings, corresponding to pleasure sensations, in a brain that is similar or identical to yours; you are thinking about whether \textit{you} will experience those pleasure sensations.

This is also true, I suspect, of those philosophers who officially refute transtemporal identity.  This is easily demonstrated by proponents of the Many Worlds Interpretation (``MWI") of quantum mechanics, in which a similar identity problem notably arises in the question of what an observer experiences at some quantum mechanical ``branching" event (cf. \cite{Wallace}).  Imagine an observer, originally in state $\ket{\Psi_{obs}}_{ready}$, who is about to perform a measurement on a quantum mechanical system in initial pure state $\ket{\Psi_{s}}$, which is a superposition of orthogonal (or mutually exclusive) states $\ket{A}$ and $\ket{B}$ such that a measurement of it will yield outcome A with 99\% probability or outcome B with 1\% probability:
\begin{equation}
\ket{\Psi_{s}} = \Bigl(\sqrt{0.99} \ket{A}+\sqrt{0.01} \ket{B} \Bigr)
\end{equation}
The measurement apparatus, originally in state $\ket{\Psi_{m}}_{ready}$, is designed so that if outcome A obtains, the observer will experience intense pain, while if B obtains, he will experience neither pain nor pleasure.  According to MWI, which assumes the universality of quantum mechanics at any scale, the experiment causes the entire system to linearly evolve according to:
\begin{eqnarray}
&& \ket{\Psi_{obs}}_{ready}\ket{\Psi_{m}}_{ready}\ket{\Psi_s} \nonumber\\
&\rightarrow& \ket{\Psi_{obs}}_{ready}  \Bigl( \sqrt{0.99} \ket{\Psi_{m}}_{A} \ket{A}+\sqrt{0.01} \ket{\Psi_{m}}_{B} \ket{B} \Bigr) \nonumber\\
&\rightarrow& \Bigl(  \sqrt{0.99} \ket{\Psi_{obs}}_{A} \ket{\Psi_{m}}_{A} \ket{A} \nonumber\\
&&\qquad\qquad\qquad+\sqrt{0.01}  \ket{\Psi_{obs}}_{B}  \ket{\Psi_{m}}_{B} \ket{B} \Bigr)
\end{eqnarray}
Notice that the final evolution in Eq. 2 remains a superposition in which a state of the observer corresponding to observation of outcome A (and thus the experience of pain) is correlated to outcome A, and vice versa for outcome B.  In collapse interpretations of quantum mechanics, the wave state in Eq. 2 instantaneously reduces to either $\ket{\Psi_{obs}}_{A}\ket{\Psi_{m}}_{A}\ket{A}$, a world in which the observer experiences intense pain with a likelihood of 99\%, or $\ket{\Psi_{obs}}_{B}\ket{\Psi_{m}}_{B}\ket{B}$, a world in which the observer avoids pain with a probability of 1\%.  Transtemporal identity is no problem in this case because the pre-measurement observer can expect to \textit{become} one or the other of the post-measurement observers.  However, because MWI is a non-collapse interpretation in which every term in a quantum wave state, no matter how macroscopic, corresponds to an actual world, MWI implies that after the measurement, there is a real observer correlated to outcome A and a real observer correlated to outcome B.  Identity fissions, and neither post-measurement observer has a better claim to ``being" the pre-measurement observer than the other \cite{Lewis}: ``All your future selves, on all your branches, are equally real and equally yours."  Like many philosophers, the MWI proponent can, in an academic setting, disavow any belief in a transtemporal identity.  Perhaps he will try to console himself by noting that the experiment will cause his identity to fission so that, after the measurement, there will exist an ``equally real" version of himself that has experienced no pain.  But I suspect that, despite fanciful arguments discounting notions of enduring identity, he will be more than a little nervous to carry out the above experiment.  I suspect that when push comes to shove, he will in fact be afraid of the pain that he -- not someone who looks and acts just like him, but \textit{he} -- is likely to experience.  

\subsection{Much Ado About Nothing}
\label{sec:nothing}

The most likely objections to the arguments in this paper will target transtemporal identity.  But even if I haven't convinced you that science without identity is unscientific, or that philosophy without identity is inapplicable to this paper, or that belief in transtemporal identity is essentially universal, it doesn't matter because the issue of identity is a red herring.

My goal in this paper is to determine whether or not conscious states can be copied, where a particular conscious state is identified with a particular person.  That is, I want to answer questions like, ``Is it possible in principle to upload myself to a computer?" or ``Is it possible to teleport me to a distant galaxy?"  To do so, I simply assume that a person's conscious state (i.e., a conscious state corresponding to someone's transtemporal identity) supervenes on some underlying physical state that can be copied, and then show that this assumption leads to a contradiction.  But if a person's conscious state does \textit{not} supervene on physical state (whether because identity is an illusion, or identity fissions, or whatever), then we come to the same conclusion that conscious states cannot be copied, because there is nothing physical that can be copied that will produce a conscious state of \textit{that person}.

Stated bluntly: if there is something physical that can be copied that causes me to be uploaded to a computer, then what are the logical consequences?  (That is a question I will explore in Section III.)  But if there is \textit{not} something physical that can be copied that causes \textit{me} to be uploaded to a computer, then we have the answer: (my) conscious states cannot be copied.

\section{WHY CONSCIOUS STATES CANNOT BE COPIED}
\label{sec:copy}
\subsection{Assumptions}
\label{sec:assumptions}

In this section, I'll lay out the structure of my argumentation as well as some of the assumptions I make.  I will let conscious state (or mental state) include identity so that two or more different people cannot experience the same conscious state.  However, different conscious states could clearly be experienced by the same person, such as if a person experiences conscious state $C_1$ and then later state $C_2$, which is temporally downstream from state $C_1$.  Also, I use ``person" to refer to any conscious entity, whether instantiated in human or other biological or even nonbiological form.

\textbf{\textit{A scientist claims that he is able to upload me to a digital computer.  Do we have reason to doubt his claim?}}  

Let's assume that the scientist's claim is true and then follow it to its logical consequences.  First, he is claiming to upload \textit{me}.  If he is right, then when he uploads me, I will subjectively experience being conscious and aware yet embodied not by a human body but by a collection of physical objects (such as digital switches) configured as a general purpose computer executing software code.  Next, because software is simply a set of instructions that can inherently be copied and executed on any general purpose computer, what happens if he uploads me to a second computer?  Will I subjectively experience both?  What would happen if those two experiences, responding to different stimuli (or inputs), start to diverge?  It might seem easy to solve the problem by adding an \textit{ad hoc} requirement that only the first software copy can be executed, but how could Mother Nature possibly enforce this rule if the two computers are spacelike separated -- i.e., causally nonlocal -- so that the temporal ordering of events is relative to an observer?  Now, instead of software, the scientist claims that he is able to create a copy of me in a different spacetime position.  He is not merely claiming that he is able to make a reasonably good physical copy of my brain, or an exact copy of my brain, or a reasonably good physical copy of my body, or an exact copy of my body.  Rather, he is claiming to be able to copy \textit{me}, identity and all.  The same problems arise as in the case of uploading me to a computer.  I will argue that the demands of special relativity are inconsistent with the ability to freely copy a conscious state because the following two assumptions lead to a contradiction:
\begin{itemize}
\item \textbf{Supervenience of conscious states:} A conscious or mental state $C_1$ supervenes on physical state $S_1$, so that instantiation of physical state $S_1$ is sufficient to create conscious state $C_1$ of a person having an identity.  If physical state $S_1$ physically evolves over time to state $S_2$ that is sufficient to create conscious state $C_2$, then state $C_2$ will be the same person (i.e., having the same identity) as state $C_1$.
\item \textbf{Copiability of conscious states:} A copy of physical state $S_1$, which is sufficient to create conscious state $C_1$, can be instantiated elsewhere in spacetime\footnote{If state $S_1$ depends on spatiotemporal location -- i.e., if that state cannot be instantiated elsewhere in spacetime -- then the assumption of Copiability is obviously false.  To use the philosopher's lexicon, two physical systems are not ``numerically identical" if they exist at different locations in spacetime, no matter how physically similar they may be.  So if one already accepts that numerical identity is necessary for the same person to experience the same conscious state, then it follows that conscious states cannot be copied or repeated.}  in a manner that does not prevent other instantiations from evolving.
\end{itemize}

Note that the second assumption (``Copiability") depends on the first assumption (``Supervenience"); Copiability can be false on its own, but if Supervenience is false, then Copiability must also be false because there is nothing physical that can be copied to produce the same conscious state of the same person. 

The phrase ``physical state" is meant in the broadest sense possible.  Classically, the physical state of a system of matter might be fully described in phase space -- i.e., in terms of the positions and momenta of each of its constituent particles.  However, we know that the classical description of the universe is only an approximation and fails to correctly describe and predict extremely small systems, the purview of quantum mechanics.  Quantum mechanically, however, it can't be said that any particular particle in a system even \textit{has} a position or momentum until measured; and while its position can be measured to nearly arbitrary precision, quantum uncertainty guarantees a trade-off in the precision to which its momentum can be simultaneously measured.  Further, the quantum mechanical description of a particle can't be separated from that of other particles with which it is entangled, making the quantum mechanical prediction of a physical system larger than a few atoms essentially impossible.  Quantum mechanics may itself turn out to be an emergent approximation of a more fundamental ontology.

Therefore I do not intend to limit ``physical state" to any known description.  Instead, by assuming that conscious state $C_1$ supervenes on physical state $S_1$, I simply mean that there is something physical about the universe that gives rise to that conscious state.  Of course, there's no guarantee that a given physical state will create a conscious state -- probably very, very few will -- but if a conscious state exists then it depends entirely on the underlying physical state.  I also assume that a given conscious state might be created by more than one physical state.  In other words, it may be the case that conscious state $C_1$ arises from any member of some large set \{$S_1$, $S_1^{*}$, $S_1^{**}$, ...\} of underlying physical states, and that instantiating any of these physical states will produce the same conscious state $C_1$ of the same person.

Further, while the following arguments apply to a physical state of any size, they are more interesting and elucidating when applied to a small or minimal physical state $S_1$ that produces conscious state $C_1$ (although state $S_1$ need not be the absolute minimum or smallest state that produces the person's consciousness).  For instance, if it turns out that the human brain\footnote{Specifying physical state $S_1$ as ``the human brain" is quite sloppy, as we really need to specify the features (e.g., relationships among cells, molecules, electrons, etc.) that define state $S_1$ and give rise to conscious state $C_1$.  If, for example, state $S_1$ depends only on how cells act like digital switches, then $S_1$ can be specified and instantiated in the form of a digital computer.  If, however, $S_1$ depends on quantum effects, then quantum no-cloning may prevent specification of $S_1$ at all.} is sufficient to create a person's consciousness, then certainly that brain \textit{plus} a body also creates the person's consciousness.  But if, as engineers, our goal is simply to create the person's consciousness as simply and efficiently as possible, then we may as well focus on just the brain.  So when I assert that conscious state $C_1$ is created by physical state $S_1$, I don't necessarily know what kinds of information or sorts of physical attributes specify that physical state $S_1$.  I also don't know how big state $S_1$ is.  Is it the local state of certain neural connections in one's brain?  The state of one's entire brain?  One's body?  The planet?  The universe?  (Although, if it turns out as a matter of fact that the smallest physical state $S_1$ sufficient to create a conscious state $C_1$ happens to involve particles throughout the entire universe, then clearly that will prevent state $C_1$ from being copied.)  

\subsection{Spacelike Separated Copies}
\label{sec:spacelike}

I will now show how the assumptions of Supervenience and Copiability conflict with the demands of special relativity.  With reference to Fig. 1, a physical state $S_1$ on which conscious state $C_1$ supervenes exists at a Point 1 in spacetime.  In other words, a conscious person having an identity determined by state $S_1$ exists at Point 1 -- let's call her Alice and designate the person at Point 1 as $Alice_1$.  For simplicity, assume that the information specifying state $S_1$ is read and then the state is destroyed or reconfigured soon afterward so that state $C_1$ no longer exists.\footnote{This is not a requirement but does simplify the analysis.  If it turns out that the mere existence of the information necessary to produce state $S_1$ is itself adequate to produce state $C_1$, then it is impossible to destroy $S_1$ without also preventing any future instantiations of $C_1$.}   The light cone of Point 1 is shown; events outside the light cone are called \textit{spacelike} (or nonlocal) and cannot have any cause-effect relationship, temporal relationship, or information connection to the event at Point 1; events inside the light cone are called \textit{timelike} and do have a temporal relationship to Point 1.  

A copy of state $S_1$ is instantiated at Point 2 (shown with its light cone) and allowed to physically evolve to a physical state $S_2^{'}$ that is sufficient to create conscious state $C_2^{'}$.  (For the moment we will ignore what may be happening at Point 3.)  The physical evolution of state $S_1$ over time is subject to physical laws and depends on both random quantum events\footnote{Whether or not quantum mechanics has anything to do with consciousness, it is clear that conscious states can correlate to (amplified) quantum events, such as the ``click" a scientist consciously notices when using a Geiger counter to measure decays of radioisotopes.} and chaotic amplifications of initial conditions such that state $S_1$ could evolve over some time period to any one of a large set of mutually exclusive states, among which $S_2$, $S_2^{'}$, and $S_2^{''}$ are possibilities.  By Supervenience, a conscious person exists at Point 2, and that person is Alice -- let's call her $Alice_2$.  $Alice_1$ and $Alice_2$ are the same person; Alice at Point 1 is Alice at Point 2, and she will subjectively experience the evolution from $C_1$ to $C_2^{'}$ because of a temporal continuity in her identity.\footnote{To reiterate, this \textit{must} be true if Supervenience is true, and if Supervenience is not true, then conscious states cannot be copied because there is no underlying copiable physical state that preserves the person's identity, and we can skip to the end of the paper.}  

\begin{figure}[ht]
\includegraphics[width=3.2 in]{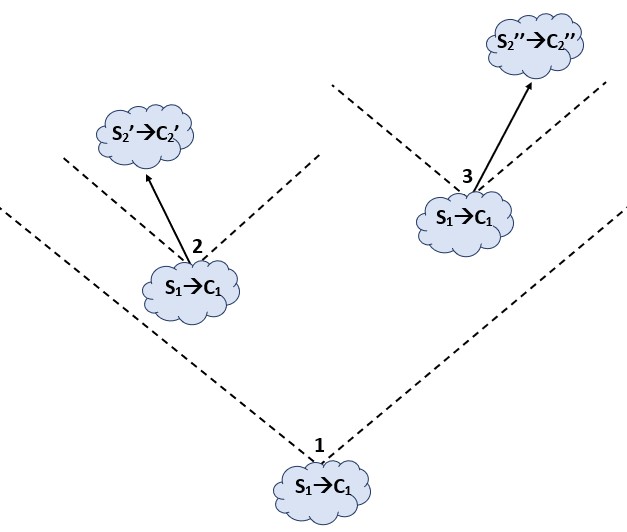}
\caption{Spacelike separated instantiations of physical state $S_1$}
\label{fig:spacelike}
\end{figure}

Fig. 1 also shows a Point 3 (with its light cone), with Points 2 and 3 lightlike to Point 1 but spacelike to each other.  A copy of state $S_1$ is instantiated at Point 3 and allowed to physically evolve to physical state $S_2^{''}$ that is sufficient to create conscious state $C_2^{''}$, where $C_2^{''}$ is consciously distinct from $C_2^{'}$ -- that is, Alice would be able to tell them apart.  (For the moment we will ignore what may be happening at Point 2.)  By Supervenience, Alice is created at Point 3 -- let's call her $Alice_3$.  Again, $Alice_3$ is the same person as $Alice_1$ and she will subjective experience the evolution from $C_1$ to $C_2^{''}$ because of a temporal continuity in her identity.  

Now, imagine for the moment in Fig. 1 that copies of state $S_1$ are instantiated at both of Points 2 and 3, but that $S_2^{'} = S_2^{''} = S_2$ and $C_2^{'} = C_2^{''} = C_2$.  Note that Points 2 and 3 are timelike to Point 1, which is a necessary condition if the information necessary to produce the copies at Points 2 and 3 depends on information collected at Point 1.  Further, Points 2 and 3 are spacelike to each other so that there is no fact about simultaneity or temporal ordering between them, nor can information pass from one to the other.\footnote{While Point 2 appears closer in time to Point 1, and thus earlier than Point 3, in fact the interval in spacetime is the metric $s=\sqrt{c^2 \Delta t^2 - \Delta x^2 - \Delta y^2 - \Delta z^2}$.  Which of Points 2 and 3 occurs earlier in time is actually relative to an observer.}  We already know that if state $S_1$ was \textit{not} instantiated at Point 2, then Alice exists at Point 3 and she will subjectively experience the evolution from $C_1$ to $C_2$ because of a temporal continuity in her identity.  However, even when $S_1$ is instantiated at Point 2, Alice exists at \textit{both} Points 2 and 3, she has precisely the same subjective experience at both points, and she will subjectively experience the evolution from $C_1$ to $C_2$.  This must be true because we already know that Alice exists in conscious state $C_1$ at Point 2.  But the creation of physical state $S_1$ at Point 3 is spacelike to creation of physical state $S_1$ at Point 2, which means that from the perspective of Point 3, there is no fact about the creation of state $S_1$ at Point 2.  The creation of $S_1$ at Point 2 can have no physical effect at Point 3.  For the same reason that Alice exists and experiences an evolution from $C_1$ to $C_2$ due to the creation of state $S_1$ at Point 2, she exists and experiences an evolution from $C_1$ to $C_2$ due to the creation of state $S_1$ at Point 3.  This may seem philosophically odd, but (so far) there is no physical contradiction.\footnote{Tappenden explains \cite{Tappenden}: ``How can two doppelgangers zillions of lightyears apart whose simultaneity we know, from Special Relativity, is entirely relative to an inertial frame, how can they share a single mind?"  The answer: no need for causal connection.}  We should be careful, however, not to imagine different Alices who happen to have very similar experiences; $Alice_1$, $Alice_2$, and $Alice_3$ are all the \textit{same} Alice.  Supervenience requires that $Alice_1$ experiences the evolution of $C_1$ to $C_2$ due to physical evolutions of instantiations of physical state $S_1$ at both Points 2 and 3.

Finally, let's return to the original configuration of Fig. 1, in which $S_2^{'} \ne S_2^{''}$ and $C_2^{'} \ne C_2^{''}$ -- i.e., that they are consciously distinct states.\footnote{We can ensure that the states are distinct by designing the systems to correlate to amplifications of cosmic microwave background radiation originating from different directions and/or points in spacetime.}  In that case, both $Alice_2$ and $Alice_3$ exist, but the underlying physical evolutions of the two physical states $S_1$ diverge.  (It makes no difference which of the two systems diverges as the divergence is relative to the other.)  For the sake of clarity, assume also that the entirety of the evolutions of states $S_1$ (at Point 2) to $S_2^{'}$ and $S_1$ (at Point 3) to $S_2^{''}$ are spacelike to each other.  Given that Alice exists at both Points 2 and 3, what does Alice actually experience in Fig. 1?  At both Points 2 and 3 she experiences state $C_1$ arising from state $S_1$, but then does she subjectively experience an evolution to $C_2^{'}$ or $C_2^{''}$?  There are exactly three logical possibilities\footnote{These three options perfectly mirror the options suggested by Saunders \cite{Saunders}: ``Nothing, both, or else just one of them?"  He concludes that only the third alternative is possible, but does not consider the effects of nonlocality.}: she experiences neither, both, or one or the other.  

Is it possible that Alice experiences neither?  Remember that Alice is created independently by instantiations at both Points 2 and 3; the instantiation evolving from Point 3 cannot be affected by the instantiation evolving from Point 2.  If Alice is created by the instantiation at Point 3 and experiences an evolution from $C_1$ to $C_2^{''}$, then nothing that happens in the evolution of state $S_1$ at Point 2 to $S_2^{'}$ can affect Alice's experience from $C_1$ to $C_2^{''}$.  Therefore, she cannot experience neither.

Perhaps Alice experiences both?  What would it feel like for Alice to subjectively experience consciously distinct evolutions from $C_1$ to $C_2^{'}$ and $C_1$ to $C_2^{''}$?  Or, more colloquially, what would it feel like for her to experience diverging streams of consciousness?  Before we indulge in speculations about higher planes of consciousness or fissioning identities, we should note the problem with locality.  Note that the experience of $Alice_2$ in Fig. 1 depends on the underlying physical evolution of $S_1$ to $S_2^{'}$, while the experience of $Alice_3$ depends on the underlying physical evolution of $S_1$ to $S_2^{''}$.  $Alice_3$ cannot experience an evolution of conscious states that depends on a physical evolution of $S_1$ to $S_2^{'}$, because $S_2^{'}$ is spacelike to $S_2^{''}$, just as $Alice_2$ cannot experience an evolution of conscious states that depends on a physical evolution of $S_1$ to $S_2^{''}$.  Therefore, Alice cannot experience an evolution of conscious state from $C_1$ to \textit{both} $C_2^{'}$ and $C_2^{''}$.

Finally, is it possible that Alice experiences one or the other?  The problem here is that because the underlying physical systems are spacelike separated, there is no means or criteria by which nature can independently select one instantiation or the other for the continuity of Alice's identity.  For instance, imagine that Mother Nature simply required that the ``first" instantiation of state $S_1$ would assume Alice's identity; that won't work because Points 2 and 3 are spacelike to each other so there is no observer-independent fact about which happens first.  Perhaps Mother Nature could choose the point having the shortest spacetime distance to Point 1, which in Fig. 1 would be Point 2.  However, again, the spacelike separation of the points prevents the passage of any signal from Point 2 to Point 3 to ``stop" the creation of Alice's conscious state $C_1$ at Point 3.

Therefore, having ruled out the only three possibilities, the copiability of conscious states to spacelike locations in spacetime leads to a contradiction.  One might object that the above argument depends on the instantiation and evolution of spacelike-separated copies of state $S_1$; perhaps there is no problem when the copies are instantiated locally to each other.  This objection fails.  If there is something fundamentally impossible about the spacelike instantiation of copies, then it must be found in their instantiation, not in their relative locations or the extent to which they are separated.  Further, the following section will show that timelike instantiations of copies of state $S_1$ are equally problematic.

\subsection{Timelike Separated Copies}
\label{sec:timelike}

The spacelike instantiations in Section B are problematic because of a lack of causal connection.  I will now show that a similar contradiction arises when we consider two timelike instantiations of state $S_1$ because they would require impermissible backward-in-time causation.

Fig. 2 is comparable to Fig. 1, the primary difference being that Point 3 is timelike to (and occurs after) Point 2, which is timelike to (and occurs after) Point 1.  For the same reasons, if we ignore what may be happening at Point 3, then $Alice_2$, who is the same person as $Alice_1$ at Point 1, exists at Point 2 and will subjectively experience the evolution from $C_1$ to $C_2^{'}$.  And if we ignore what may be happening at Point 2, then $Alice_3$, who is the same person as $Alice_1$ at Point 1, exists at Point 3 and will subjectively experience the evolution from $C_1$ to $C_2^{''}$.  

\begin{figure}[ht]
\includegraphics[width=3.2 in]{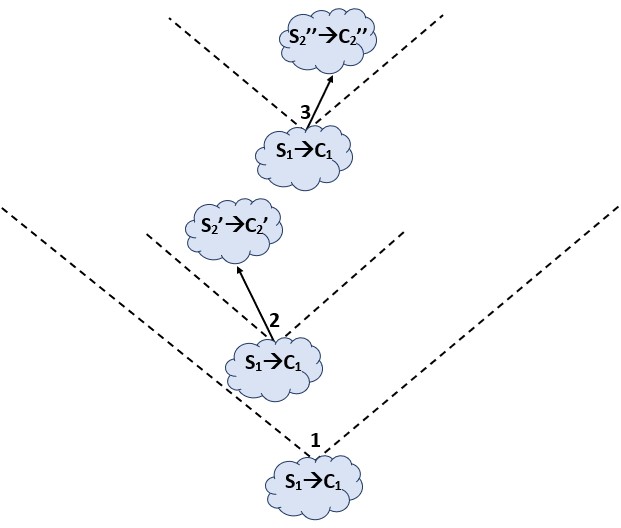}
\caption{Timelike separated instantiations of physical state $S_1$}
\label{fig:timelike}
\end{figure}

Now, imagine for the moment in Fig. 2 that copies of physical state $S_1$ are instantiated at both of Points 2 and 3, but that $S_2^{'} = S_2^{''} = S_2$ and $C_2^{'} = C_2^{''} = C_2$.  By the assumption of Supervenience, Alice exists and has precisely the same subjective experience at \textit{both} Points 2 and 3 and will subjectively experience an evolution from $C_1$ to $C_2$.  To the extent that any conscious state, when it is experienced, is subjectively experienced as occurring ``now," states $C_1$ at Points 2 and 3 must be experienced identically and thus must be subjectively experienced simultaneously.  So, even though Point 3 objectively occurs after Point 2, Alice's conscious experience (which includes the subjective experience of ``now") is the same at both points.  After all, if they were subjectively experienced as different times, such different experiences would manifest themselves in different conscious states.  While this may seem philosophically odd, again there is no contradiction.\footnote{Zuboff explains \cite{Zuboff}: ``This experience [across brains] of being you, here, now, would be numerically the same whenever, as well as wherever, it was realized."}

Unlike in Section B in which the copies of state $S_1$ are spacelike separated, $Alice_3$ is in fact created after $Alice_2$, in which case a causal connection is possible in principle.  However, state $S_1$ is already assumed (by Supervenience) to be sufficient to create Alice in conscious state $C_1$, so the instantiation of state $S_1$ at Point 3 must necessarily create Alice in state $C_1$, independently of her creation at Point 2 (or any other event prior to Point 3).  In other words, the assumptions of Supervenience and Copiability prevent the kind of causal connection that would invalidate these arguments.  For example, consider the case in which creation of $Alice_2$ causally prevents the creation of $Alice_3$; if that were true, then creation of conscious state $C_1$ at Point 3 would depend on facts that transcend physical state $S_1$, thus contradicting Supervenience.   Therefore, if state $S_1$ is sufficient to create Alice in state $C_1$, then instantiation of $S_1$ at Point 3, and its corresponding creation of $Alice_3$, cannot be altered by the existence of $Alice_2$.  For the same reason that Alice exists and experiences an evolution from $C_1$ to $C_2$ due to the creation of state $S_1$ at Point 2, she also exists and experiences an evolution from $C_1$ to $C_2$ due to the creation of state $S_1$ at Point 3.  Again, we should be careful not to imagine different Alices who happen to have very similar experiences; it is the same Alice. 
 
Finally, let's return to the original configuration of Fig. 2, in which $S_2^{'} \ne S_2^{''}$ and $C_2^{'} \ne C_2^{''}$ (i.e., they are consciously distinct states).  In that case, both $Alice_2$ and $Alice_3$ exist, but the underlying physical evolutions of the two states $S_1$ diverge.  Given that Alice exists at both Points 2 and 3, what does Alice actually experience in Fig. 2?  At both Points 2 and 3 she experiences state $C_1$ arising from state $S_1$, but then does she subjectively experience an evolution to $C_2^{'}$ or $C_2^{''}$?  Like in the case of spacelike separated copies of state $S_1$, there are exactly three logical possibilities: she experiences neither, both, or one or the other.  

Regarding neither, I have explained why instantiation of state $S_1$ at Point 3, and its evolution to state $S_2^{''}$, will produce Alice subjectively experiencing state $C_1$ to $C_2^{''}$ independently of instantiation of state $S_1$ at Point 2.  Therefore, it cannot be neither.

Regarding both, the problem here is that Alice experiences the evolutions of $C_1$ to $C_2^{'}$ and $C_1$ to $C_2^{''}$ as simultaneous (from her subjective perspective), because they both begin at the same conscious state $C_1$.  Therefore, for Alice to experience both conscious evolutions, it must be the case that the conscious experience of $Alice_2$ depends not only on the underlying physical evolution of $S_1$ (at Point 2) to $S_2^{'}$, but also on the underlying physical evolution of $S_1$ (at Point 3) to $S_2^{''}$.  However, if Alice is to experience conscious states in real time with their underlying physical states, this scenario would require impermissible backward-in-time causation, because what she experiences prior to Point 3 would depend at least in part on events occurring \textit{after} Point 3.  Further, because of physical indeterminism (e.g., chaos and quantum mechanical randomness), it cannot be known at Point 2 whether, or when, additional copies of state $S_1$ will be instantiated.  Therefore, it would be impossible for Alice to experience both an evolution of $C_1$ to $C_2^{'}$ and $C_1$ to $C_2^{''}$ if she experiences them in real time with the underlying physical evolutions.  

The only possible way for Alice to experience both evolutions would require two facts: first, that Alice's conscious states are experienced \textit{after} (i.e., not in real time with) the physical states that produced them; and second, that Alice could not experience \textit{any} conscious states until the instantiation of additional copies of state $S_1$ is no longer possible.  In other words, Alice would only be able to experience conscious state $C_1$ created by multiple copies of state $S_1$ \textit{after} the final copy of state $S_1$ was instantiated.  Thus, Alice can only subjectively experience consciousness after which time her conscious states can no longer be copied or repeated -- in which case Copiability is false. 
 
Finally, regarding whether Alice will experience one evolution or the other, I have explained why instantiation of state $S_1$ at Point 3, and its evolution to state $S_2^{''}$, will produce $Alice_3$ subjectively experiencing state $C_1$ to $C_2^{''}$.  Therefore, if Alice is to experience conscious states in real time with their underlying physical states, whether or not $Alice_2$ will experience the evolution from $C_1$ to $C_2^{'}$ depends on the \textit{future} fact of whether $Alice_3$ is created by instantiation of state $S_1$ at Point 3, another impermissible backward-in-time causation.  Thus, for Alice to experience one evolution or the other, not only would Alice's conscious states necessarily be experienced long after the physical states that produced them have long since disappeared, but Alice could not experience any conscious states until the instantiation of additional copies of state $S_1$ is no longer possible.  Thus, Alice can only subjectively experience consciousness on or after which her conscious states can no longer be copied or repeated -- in which case Copiability is false.  

Therefore, having ruled out the only three possibilities, the copiability of conscious states to timelike locations in spacetime leads to a contradiction.  In conjunction with the conclusion of Section B, it follows that Copiability is false.

\section{DISCUSSION}
\label{sec:discussion}

The assumptions of Supervenience and Copyability lead to a contradiction, whether copies of a conscious state are created spacelike or timelike.  Therefore, Copyability, which depends on Supervenience, is false.  Consequently, a conscious state cannot be copied or repeated (unless something about the universe prevents the copies from evolving differently); said differently, if two instances of the same conscious state could evolve to different conscious states, they cannot both exist.  This conclusion does not speak to the veracity of the statement, ``Consciousness is created by the brain," but if that statement is true, then there must be something physical about the brain that prevents it from being copied.  Why should this be the case?  And what are some potential implications?

\subsection{Explanatory Hypotheses}
\label{sec:hypotheses}

Why might it be fundamentally physically impossible to copy or repeat conscious states?  It may be the case that there is no way to adequately measure physical state $S_1$, or if there is, that there is no way to recreate it elsewhere in spacetime.  Some potential reasons include: the information necessary to specify the physical state is contained in distant particles that cannot be reached; attempting to measure the information inherently destroys it, for example, because of reduction of a quantum wave function; the information cannot be measured with adequate precision due, for example, to quantum uncertainty; the state includes quantum entanglements with distant particles that cannot be recreated; and/or the state is too large to measure or instantiate elsewhere.

The assumption of Supervenience does not depend on physical state $S_1$ being a quantum state, nor do the arguments in this paper depend on any relationship between consciousness and quantum mechanics.  Having said that, many of the above explanatory hypotheses for the physical inability to duplicate conscious states invoke quantum mechanics.  This should not be entirely surprising; the only known physical mechanism that prevents the instantiation of multiple copies of the same entity is quantum no-cloning, a no-go theorem that asserts the impossibility of creating an identical copy of an unknown quantum state.  It may well be the case that whatever physical state $S_1$ is required to create conscious state $C_1$ depends on quantum information that prevents it from being copied.  Aaronson \cite{Aaronson}, for example, points out that if a conscious chunk of matter is ``unclonable for fundamental physical reasons," then that unclonability could be a consequence of quantum no-cloning if the granularity a brain would need to be simulated at in order to duplicate someone's subjective identity was down to the quantum level.  

Further, the non-repeatability of a conscious state implies irreversibility, and the concept of irreversibility pervades physics: classically in the form of the Second Law of Thermodynamics and increasing entropy\footnote{Strictly speaking, classical physics is time-reversible; increasing entropy is entirely a statistical result and is not an immutable law of nature.}, and quantum mechanically in the form of decoherence \cite{Haroche}, which occurs ``as soon as a single [correlated] quantum is lost to the environment."  Since there will always be those who claim that any quantum mechanical process is reversible in principle by acting on the system with the reverse Hamiltonian, we could limit ourselves to ``any fact of which the news is already propagating outward at the speed of light, so that the information can never, even in principle, be gathered together again in order to 'uncause' the fact" \cite{Aaronson}.  If we indeed live in a universe with nonnegative curvature, as is currently believed, and given that almost the entire night sky is black, we can regard virtually every photon emitted into space as irretrievable.  Thus if it turns out that consciousness can only result from irreversible processes, that irreversibility may itself be manifested in a physical state that cannot be read or copied because some of the information specifying it is embedded in photons streaming through space at the speed of light.

\subsection{Implications}
\label{sec:implications}

First, of course, the conclusion that conscious states cannot be copied or repeated instantly renders moot the philosophical problems mentioned in the Introduction.  Two others will be broached here.

\textbf{Consciousness is not algorithmic.}  The possibilities of mind uploading and computer consciousness depend on whether consciousness is fundamentally algorithmic -- that is, whether consciousness can be reduced to a finite set of input-dependent instructions.  Because an algorithm can be executed in the form of software on any general-purpose computer independently of its underlying physical substrate, any algorithm can be copied with no fundamental limitation on the number of copies that can be executed; it can also be repeated on the same computer by resetting that computer and then executing the software again.  If a conscious state cannot be copied or repeated, but an algorithm can, then an algorithm cannot produce a conscious state.  That inevitably leads to the conclusions that Strong Artificial Intelligence (``Strong AI") is likely false (cf. \cite{Searle}) and that mind uploading and computer consciousness are likely impossible.

\textbf{Wigner's Friend is not conscious.}  In Eq. 1, we were told that the quantum mechanical system $\ket{\Psi_{s}}$ is initially in a superposition of states $\ket{A}$ and $\ket{B}$; when the observer performs a thousand experiments on systems identically prepared in state $\ket{\Psi_{s}}$, he finds that around 99\% of them are measured in state $\ket{A}$ and the remaining in state $\ket{B}$.  However, what if the observer had been fooled?  What if instead of a thousand systems in a pure state, he was actually provided with 990 systems already in state $\ket{A}$ and 10 already in state $\ket{B}$ -- i.e., a mixed state?  Is there a way for him to modify the experiment to figure this out?  Yes: by making measurements in a new basis in which one element, $\ket{C}$, is parallel to $\ket{\Psi_{s}}$ and the other element, $\ket{D}$, is perpendicular.  If each system was originally in a pure state, then all 1000 will, with certainty, be measured in state $\ket{C}$; if the systems were mixed, then some are likely to be measured in state $\ket{D}$.\footnote{This is because the systems starting in state $\ket{A}$ are in a superposition of $\ket{C}$ and $\ket{D}$ with nonzero amplitudes, so when each such system is measured in the \{$\ket{C}$,$\ket{D}$\} basis, some will be measured as $\ket{C}$ and some as $\ket{D}$; and vice versa for those systems starting in state $\ket{B}$.  However, because there is some nonzero probability that all 990 initially $\ket{A}$ states and all 10 initially $\ket{B}$ states are, by random quantum ``accident," measured in state $\ket{C}$, which is identical to the outcome predicted for systems in a pure state, it actually requires infinitely many interference experiments to confirm a pure state (cf. \cite{Okon}).}  While the principle is relatively easy to explain, the actual process of measuring in the new basis -- called an interference experiment -- is prohibitively complex in all but the simplest situations.

In Eq. 2, let's rename the observer either Schrodinger's Cat (``SC") or Wigner's Friend (``WF"); both refer to essentially identical thought experiments in which an isolated macroscopic system involving some quantum mechanical event is allowed to evolve until finally observed by a Super-observer (Schrodinger or Wigner, respectively).  When Wigner finally confers with his friend post-measurement, WF of course reports having measured either outcome A or B.  But did WF collapse the wave function when he performed the measurement?  Or did WF remain in the superposition shown in Eq. 2 until Wigner conferred with his friend, in which case Wigner's observation of WF collapsed the wave function?  This is the same question as whether WF was in a pure or mixed state at the time of Wigner's observation, and the only way for Wigner to know is by repeatedly doing an interference experiment on WF in a basis parallel to the state in Eq. 2.  But the only way to do \textit{that} would be to repeatedly measure WF in a coherent superposition of states, which requires that WF's state (specifying his body, brain, memories, etc.) is itself reversible.  Such manipulations, if they were possible, would require erasing WF's memory and further require WF to experience ``the same mental processes over and over, forwards in time as well as backwards in time" \cite{Aaronson}.  Deutsch \cite{Deutsch} describes a hypothetical method for experimenting on WF by having him send a message to Wigner that is uncorrelated to the measurement outcome\footnote{But see \cite{Salom}, which points out why a post-measurement message from WF will always be correlated to the outcome.}, but concedes that such an experiment requires memory erasure.  While many scholars have asserted the ``in-principle" possibility of such an experiment\footnote{But see \cite{Barros}, which argues that the thermal isolation required for such reversibility precludes performing the experiment on any living thing we might regard as conscious.}, Deutsch attempts to overcome otherwise insurmountable technical hurdles of performing a WF-type experiment on humans by instead positing conscious \textit{computers}.

However, if a conscious state cannot be repeated, as I showed in Section III, then this kind of manipulation on a conscious observer is not possible.  In other words, if the WF (or SC) experiment can actually be performed, then WF (or SC) is not conscious; if he is conscious, then the experiment cannot be performed.  Further, if consciousness is not algorithmic, then Deutsch's attempt to salvage his experiment by substituting a computer for a person fails.  Because the ability to measure WF in a coherent superposition requires reversibility, and because a conscious state is not reversible (because it cannot be repeated), it follows that WF's conscious measurement is related to an irreversibility in the measurement process.  A conscious state can only arise after reversal of the state is no longer possible -- that is, when the state is truly irreversible.  Therefore, consciousness may be a direct result, and indication, of the irreversibility of a physical process.\footnote{Aaronson comes to a similar conclusion \cite{Aaronson} and notes that, if correct, it may have interesting cosmological implications, such as the universe having nonnegative curvature.}     

\subsection{Concluding Remarks}
\label{sec:conclusion}

\begin{quote}
\emph{Conscious states are produced by the brain; the brain is a just a chuck of matter; it is copiable in principle; therefore, copying a conscious state is merely a technological problem.}
\end{quote}

The allure of this faulty logic is irresistible and has led physicists and philosophers into a minefield of intractable problems and seeming paradoxes.  More careful scrutiny of the assumptions, particularly in light of physical limitations imposed by special relativity on copies of conscious states instantiated at different points in spacetime, yields a contradiction: there is something about consciousness that prevents it from being copied or repeated at will.

\begin{acknowledgments}

I would like to thank Martin Bier, Brian Pitts, Scott Aaronson, Paul Tappenden, Igor Salom, Kenneth Augustyn, Gary Ihas, Christopher Lind, and Annie Knight for their guidance, comments, and stimulating conversations.  This research did not receive any specific grant from funding agencies in the public, commercial, or not-for-profit sectors.

\end{acknowledgments}

\end{document}